\documentclass{elsarticle}
\newcommand{\bvec}[1]{ {\mathbf #1} }

\usepackage{hyperref}
\journal{New Astronomy}

\biboptions{authoryear}

\begin{document}

\begin{frontmatter}

\title{Automatic detection of white-light flare kernels in SDO/HMI intensitygrams}
% \tnotetext[mytitlenote]{Fully documented templates are available in the elsarticle package on \href{http://www.ctan.org/tex-archive/macros/latex/contrib/elsarticle}{CTAN}.}

\author[auuk]{Lucia Mravcov\'a}
\ead{lucka3@gmail.com}

\author[asu,auuk]{Michal \v{S}vanda \corref{mycorrespondingauthor}}
\cortext[mycorrespondingauthor]{Corresponding author}
\ead{michal@astronomie.cz}

\address[auuk]{Astronomical Institute, Charles University in Prague, Faculty of Mathematics and Physics, V Hole\v{s}ovi\v{c}k\'ach 2, CZ-18000 Prague 8, Czech Republic}
\address[asu]{Astronomical Institute (v. v. i.), Czech Academy of Sciences, Fri\v{c}ova 298, CZ-25165 Ond\v{r}ejov, Czech Republic}

\begin{abstract}
Solar flares with a broadband emission in the white-light range of the electromagnetic spectrum belong to most enigmatic phenomena on the Sun. The origin of the white-light emission is not entirely understood. We aim to systematically study the visible-light emission connected to solar flares in SDO/HMI observations. We developed a code for automatic detection of kernels of flares with HMI intensity brightenings and study properties of detected candidates. The code was tuned and tested and with a little effort, it could be applied to any suitable data set. By studying a few flare examples, we found indication that HMI intensity brightening might be an artefact of the simplified procedure used to compute HMI observables. 
\end{abstract}

\begin{keyword}
Sun: activity\sep Sun: flares
\end{keyword}

\end{frontmatter}

% \linenumbers

\section{White-light flares on the Sun}

The Sun is considered a prototype star, where, compared to the other stars, we have a luxury to study it in detail, thanks to the availability of high-cadence high-resolution observations together with long-term synoptic recordings. On top of that, the Sun is an active star, where phenomena like sunspots, faculae, prominences, and flares are connected to the variability of magnetic fields at various levels of solar body and its atmosphere. The flares in particular are considered the most violent active phenomena, with a direct influence on the neighbourhood of the Earth, thereby being responsible for space weather events. 

The flares on the Sun are a consequence of the reconnection of the entangled magnetic field above active regions \citep{2011LRSP....8....6S}. The phenomenon of a `flare' consists of effects connected to intensive atmospheric heating, formation of particle beams, and release of coronal mass ejections. Flares on the Sun depict various energies with smaller ones occurring frequently whereas the large flares are rare \citep[e.g.][]{1993SoPh..143..275C,1995PASJ...47..251S,2000ApJ...535.1047A}. Solar flares are observed at all wavelengths from radio waves to gamma-rays. However, they are mostly prominent at radio wavelengths due to the non-thermal processes in the plasma and at high-energy (EUV and X-rays) range of wavelengths due to the high temperature at a reconnection site \citep{2008LRSP....5....1B}. 

In the range of the visible light (VL), the flares can easily be observed in chromospheric spectral lines, such as H$\alpha$. It is believed that the chromospheric-line emission is stimulated by the collisional excitation of the chromosphere under the reconnection site by electron beams that are accelerated during the reconnection. 

Despite the fact that most of our information about solar flares was devised from analysis of observations obtained either in chromospheric emission lines or radio or EUV emission, the first ever observed solar flare was seen by a naked eye in the white light \citep{1859MNRAS..20...13C,1859MNRAS..20...15H}. 

A class of white-light flares (WLF) is considered something special, where the true origin of the continuum emission is not entirely clear \citep[see a recent overview by][]{2016SoPh..291.1273H}. The original idea was based on an assumption that in the strong flares, the electron beams penetrate all the way down into the photosphere, where they by collisions heat the photospheric plasma, which then radiates thermally. Recent simulations \citep[such as by][]{2016AN....337.1020M} do not support this idea, because it seems that most of the energy in beams is deposited already at chromospheric levels and only a small fraction may propagate further down. \cite{2008ApJ...675.1645F} suggested an alternative model in which the energy is transported from the reconnection site to the very deep layers of solar atmosphere by Alv\'en-wave pulses, which would provide an environment for the electrons in the lower atmosphere to be efficiently accelerated to high energies. Those accelerated electrons then collisionally stimulate the white-light (WL) emission.

It was found that the WL emission correlates well with hard X-ray sources \citep[e.g.][]{2003A&A...409.1107M,2006SoPh..234...79H,2011ApJ...739...96K,2012ApJ...753L..26M}, which would indicate that the source region of the WL emission is in the low chromosphere. In case of a 30 Sep 2002 flare the kernels of a continuum emission even moved cospatially with the X-ray footpoints \citep{2006ApJ...641.1217C} indicating that the source term for the emission must be the collisions by non-thermal electrons. \cite{2016RAA....16..177H} demonstrated by studying a set of 13 WLFs that stronger WL emission tends to be associated with a larger population of high energy electrons. A similar study by \cite{2016ApJ...816....6K} indicated that for the WL production, flare-accelerated electrons of energy $\sim50$~keV are the main source. \cite{2007ApJ...656.1187F} on the contrary pointed out that the bulk of the energy required to power the WLF resides at lower energies of around 25~keV. A thorough study of X1 flare \citep{2017ApJ...836...12K} showed that the emission may originate from two layers in the atmosphere, the dense chromospheric condensation with a low optical depth and stationary beam-heated layers just below this consensation. A very strong dependence of the WL emission on the local atmospheric conditions and thus ``penetrability'' of the atmosphere by flare-accelerated electron from the reconnection site was indicated also earlier by \cite{2005ApJ...618..537C}.

\cite{2006ApJ...641.1210X} by using multiwavelength high-resolution observations showed that the emission energy budget cannot be ballanced if only a direct heating of the emitting regions is considered. An alternative explanation for the white-light emission observed during WLF is that it is a combination of the spectral continuum (Paschen and Balmer continuum), which forms in the chromosphere, and a photospheric emission, which is stimulated not by beams, but by radiative heating from the chromosphere. This so called `backwarming' model was devised by \cite{1989SoPh..124..303M}. It seems that for at least some flares classified as WLFs, mostly the strong ones, the presence of a radiative backwarming is necessary to explain the observed VL intensities \citep[e.g.][]{2003A&A...403.1151D, 2003ApJ...595..483M,2010ApJ...711..185C}. 

\cite{2008ApJ...688L.119J} reported on a very fine structure of white-light emission, where the points of brightening were not larger than 300~km, much beyond the resolution of synoptic telescopes. The authors concluded that with a sufficient resolution, every flare must have a white-light emission, in favour of the backwarming model.

It has to be noted that indications for flares in the white light were also observed on other stars, first on young or magnetically interacting stars, such as the RS CVn type, and magnetically active M dwarfs.  Flares on Sun-like stars were not seen until the availability of the high-cadence high-precision photometry \citep{2000ApJ...529.1026S}. Recently \cite{2012Natur.485..478M} reported on observations of large-energy flares on Sun-like stars recorded in the {\it Kepler} light curves. In general, it is believed that the most important contribution to flares observed in the range of the visible light on other stars comes from the Paschen continuum \citep[e.g.][]{2006ApJ...644..484A}. The appearance of the flares is not reserved to only stars of a late spectral type, but apparently also to hotter stars of type A \citep[e.g.][]{2012MNRAS.423.3420B,2016ApJ...831....9S}, however recent observations do no confirm this idea \citep{2017MNRAS.466.3060P}.

The aim of our study is twofold. First, we develop a code that allows a detection of brightenings in the intensity images that have properties of WLF kernels. We describe the algorithm behind our code, properly determine the values of tunable parameters and test it. Second, we show a rather simplistic application to the results returned by the code by examining simple properties of our sample of detected WLFs. The aim of the second goal is to show that the code returns data product suitable for a further investigation of the physics of WLFs. 

In a future we believe that the systematic research of WLFs from synoptic instruments such as SDO/HMI may answer questions regarding the WLF's morphology, which was studied only scarcely. For instance, some recent studies \citep[e.g.][]{1986lasf.conf..483M,1995A&AS..110...99F} indicate that there might be two classes of WLFs, ``type I'' (the brightest events showing increase in contrast towards blue end of the spectrum) and ``type II'' (with flatter spectra, kernels appear often as faint and diffuse wave-like features). \cite{1995A&AS..110...99F} claimed that these two types must have a different origin of heating, e.g. the observable properties of type II WLFs are not consistent with the heating flux coming down from the corona, but are rather consistent with a local heating in the lower atmospheric layers \citep[e.g.][]{1999ApJ...512..454D, 2001ChJAA...1..176C}. A large catalogue of WLFs from synoptic instruments is a necessary starting point for such studies. 

\section{Data}

The routine availability of synoptic solar observations by Solar Dynamics Observatory (SDO), especially from Helioseismic and Magnetic Imager \citep[HMI;][]{2012SoPh..275..229S} in the range of visible light, made this data set an ideal source to carry out a systematic search for visible-light flares. HMI continuously observes the full disc of the Sun in a FeI (617.3~nm) line, which is scanned at six positions throughout the line profile. From these six filtergrams, the pseudo-continuum images $I_{\rm c}$ (together with an estimate of the line depth and line width and also and estimate for the Doppler velocity and line-of-sight magnetogram) are reconstructed by a technique described by \cite{2012SoPh..278..217C}. The algorithm used is known to perform poorly in magnetised regions \citep[e.g.][]{2015SoPh..290..689C}, but estimates of line depth and continuum intensity seem not to suffer too much from algorithm artefacts. Let us keep the discussion about the actual origin of the emission recorded in $I_{\rm c}$ images for later and assume that $I_{\rm c}$ images can be used to systematically investigate appearance of WLFs.

The HMI instrument was used for investigation of individual solar flares recently \citep[e.g.][]{2011SoPh..269..269M}, where the authors investigated the flare appearance in all six filtergrams used to compute the line-of-sight quantities. They found e.g. a suspicious behaviour of the derived Doppler velocity in the presence of the flares, which was later re-investigated \citep{2014SoPh..289..809M}. It was found that there are obvious artefacts in HMI observables in the WLF transients due to the time-lag in sampling of the spectral line by HMI, which led to an apparent change of the shape of the spectral line and thus false values of Doppler velocity. The authors point out that similar effects may be expected in the derived intensity of the line-of-sight magnetic field. 

Thus we took and opportunity and searched for $I_{\rm c}$ emission causally connected to a set of solar flares. For the purpose of this work, we initially focused to strong flares only that were classified as M5.0 or stronger, thus with an X-ray 0.1--0.8~nm flux of $5\times10^{-5}$~W\,m$^{-2}$ or larger. The lower limit is a technical one, white-light emission was observed for flares as weak as C-class \citep[e.g.][]{2006SoPh..234...79H}, and we propose to extend our work towards weaker flares in the future. In order to avoid possible problems with projection effects, we further limited our sample to flares that ignited not farther than 50 heliographic degrees from the disc centre. Those WLF candidates were searched for in the archive of events detected by GOES satellites, the first one considered was a M6.6-class flare that occurred on February 13th 2011 (maximum at 17:28 UT), whereas the last one was a M5.6-class flare on August 24th 2015 (maximum at 07:26 UT). Between these two dates we selected 54 WLF candidates (including the two already mentioned) with a maximum X-ray class of X5.4 (March 6th 2012). The weakest WLF candidate out of 54 was classified as M1.8 (July 5th 2012) and was included in the analysis, only because it was followed by a M6.2-class flare 50 minutes later and was accidentally recorded in the same investigated datacube. 

For each flare candidate we tracked a 3-hour datacube around the time of the flare maximum by using publicly available JSOC\footnote{http://jsoc.stanford.edu} tools. The cube was tracked with a Carrington rotation rate at a full HMI resolution (pixel size of 0.5") and we focused to region centred on the flare location (as given by GOES data and verified in NASA's SolarMonitor\footnote{http://www.solarmonitor.org}) with a field-of-view of 768$\times$768~px. We retrieved not only $I_{\rm c}$ data, but also line-of-sight magnetograms that are both spatially and temporally aligned in order to perform a fast analysis of the magnetic field in WLF candidates. Such datacubes were prepared for a search of $I_{\rm c}$ brightenings connected to the flare. 

\section{Methods}

Input of the program is a datacube in FITS format where two coordinates are spatial and the third is time. The size in both space directions must be the same. There is also an optional free parameter and that is the multiple $\sigma_M$ of the standard deviation further described in the next paragraph. If it is omitted the program uses default value defined in the code which was set after analysing a number of different values. This process is also explained later in the next paragraphs.

The program reads input and stores it in a 3D array. For each spatial point we find and eliminate long-time trend in intensity. It is achieved by fitting the time-dependent intensity in every spatial point with five degree polynomial. Then we subtract this polynomial from the original values. After that we count the expected value and the standard deviation of intensity for each point in the space domain. The program then stores every point from the 3D array, where intensity is higher than the sum of the expected value and a multiple $\sigma_M$ of the standard deviation.

Now we find groups of points where a WLF could occur. These groups are found for each cut in time so now we search in 
2D arrays. It is done by breadth-first search algorithm (BFS) that looks around each point, where intensity is high enough, to see if any of its eight neighbours has also intensity above desired level. If a group of points is large enough (60 points) the program stores it and begins searching for another group until all points are processed. There is also a boundary on the maximum points (3000 points) in one group. It was added to compensate for missing frames in some datacubes. Otherwise, the frame after the missing one is detected as a sudden large-scale (too large to be a WLF) brightening, thereby leading to false-positive detection of flare kernels. 

For each group found the program counts expected value of both space coordinates and thus finds the centre of it. Then using BFS algorithm the program tries to find out if near the centre of one group is the centre of another group which takes place in the following time moment. If the time interval between the first and the last centre in the found set is long enough (limit is set to 4 frames which corresponds to 3 minutes) these groups are stored. For the same reason as in the first BFS there is a boundary (1000 points) on the maximum amount of points in one set. Now all points, that are in the stored groups, are those where a WLF should have occurred and are written in the output file.

The results of the code are sensitive to the value of $\sigma_M$, which plays a role of a free parameter. It is clear that a lesser value of the threshold will increase the number of detected points which in fact will be false positives, that is the points, where the intensity randomly exceeded this threshold. A large value will decrease the susceptibility of the code to random noise, however it will also decrease the number of darker points belonging to flare kernels. To set an optimal value of the threshold, we studied a large range of values and followed two variables.

\begin{enumerate}
\item The total number of points detected as flare kernels. We assumed that the total area of flare kernels would be much smaller than the total area of the field-of-view. Hence when the selected threshold was larger than optimal, we expected that the number of detected points would decrease slowly, as darker kernel points would fall below the increasing threshold. On the other hand, for the thresholds smaller than optimal, the number of false-positive (noise) points would increase rapidly, because they were caused by random fluctuations ($p$-modes for instance) and their location in the field-of-view was not confined to flare kernels. 
\item The spread of the location of the points around their gravity centre evaluated by standard deviation of the positions in both directions independently. The idea behind was that when only flare kernels were detected, they would be sort of confined, whereas when also random noise was detected as false positives, they would be located again all over the field-of-view and the spread would increase. 
\end{enumerate}

\begin{figure}[!t]
\includegraphics[width=0.5\textwidth]{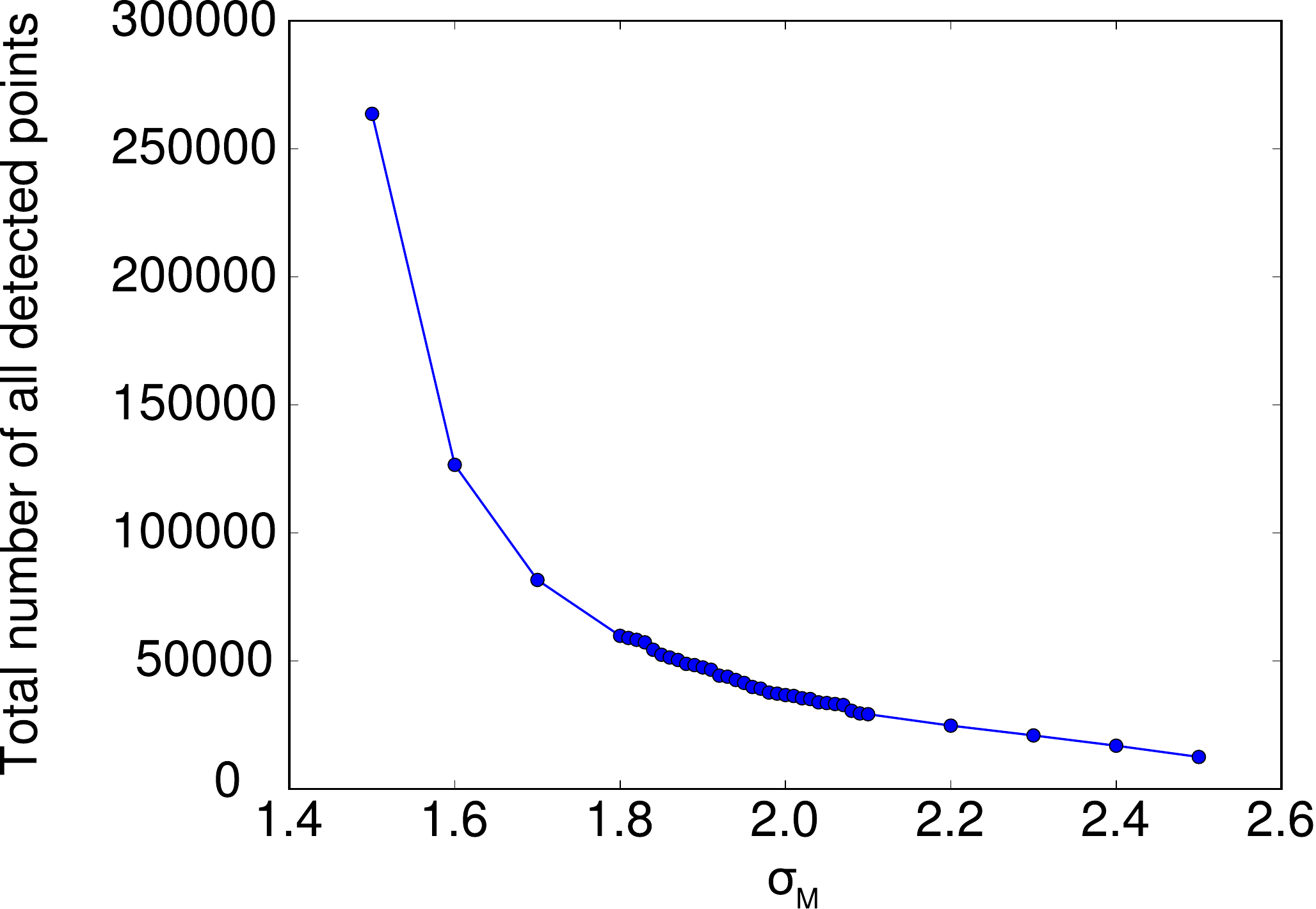}
\caption{Dependence of the total number of points detected as pixels of WLF kernels in all flares on the the value of $\sigma_M$.}
\label{fig:sigma_number}
\end{figure}

\begin{figure}[!t]
\includegraphics[width=0.5\textwidth]{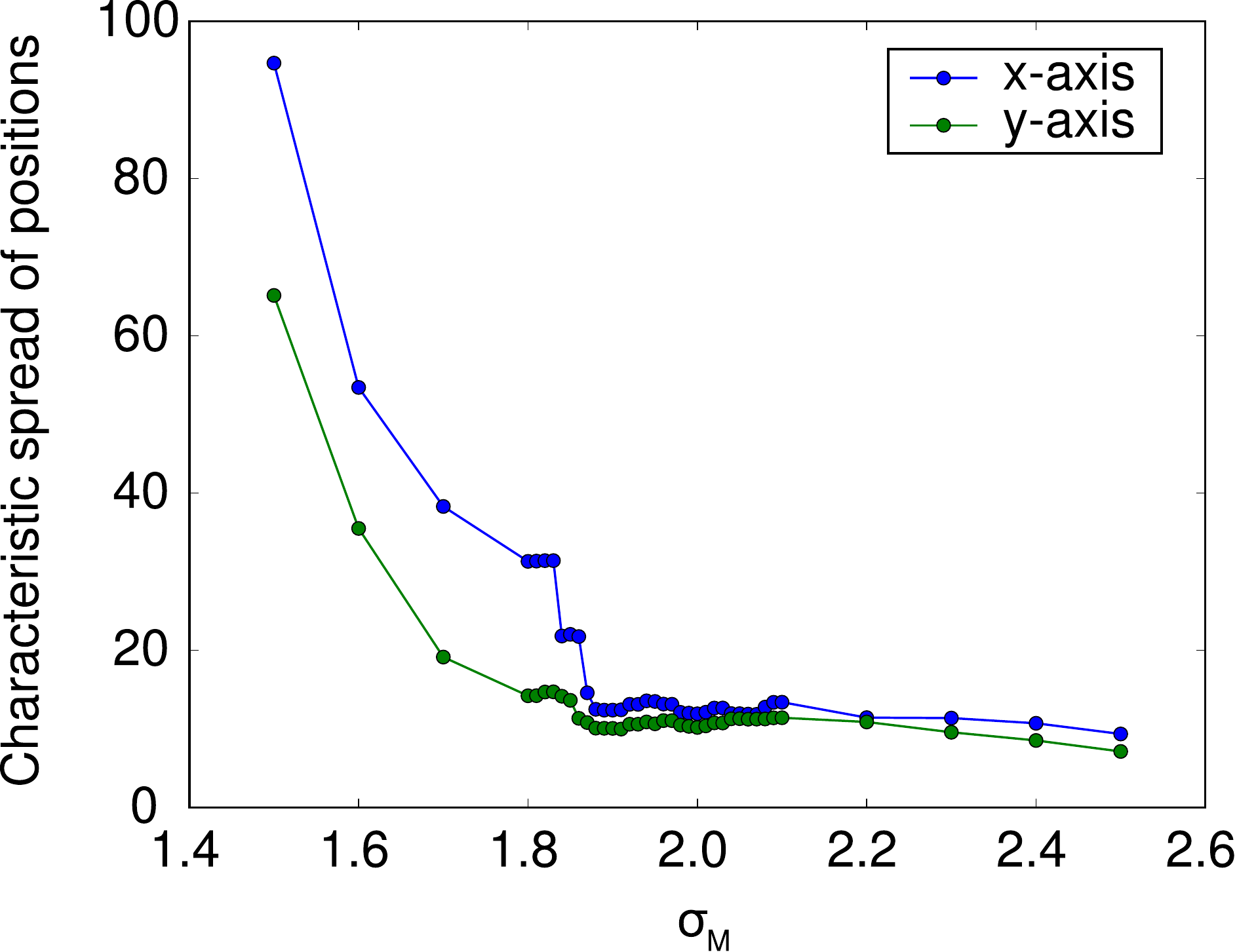}
\caption{Dependence of a characteristic spread of points detected as pixels of WLF kernels in all flares on the the value of $\sigma_M$.}
\label{fig:sigma_spread}
\end{figure}

\noindent It turned out that the proposed tests returned an unambiguous value of the optimal threshold value. As it can be seen from Fig.~\ref{fig:sigma_number} and \ref{fig:sigma_spread}, the behaviour of both selected quantities was as we expected and the optimal value for the threshold was thus 1.88. 

\section{Results}

From the 54 WLF candidates the code detected an $I_{\rm c}$ emission in 25 cases. Interestingly, there does not seem to be a clear distinction of the flares with and without $I_{\rm c}$ emission as far as the flare class is concerned, both groups span over all investigated class range. 

\begin{figure}[!t]
\includegraphics[width=0.7\textwidth]{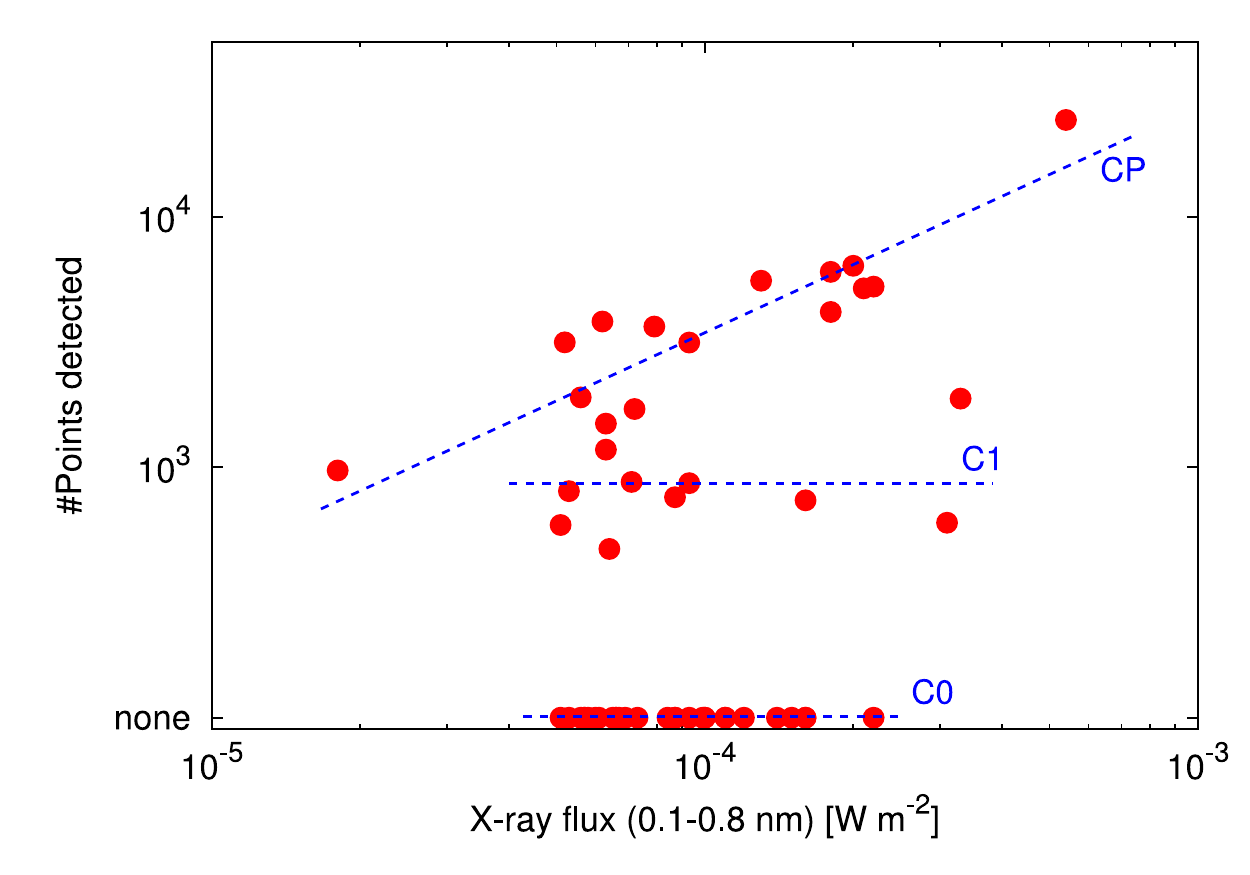}
\caption{Dependence of the total number of points in the WLF kernels on the flare strength. The dashed lines indicate the division of the flares into three classes mentioned in the text, they do not represent any fit of any kind.}
\label{fig:classes}
\end{figure}

We investigated the dependence of the detected area (integrated in time) of the WLF ribbons on the flare intensity. It seems that there apparently are three different classes: a class (C0) without detected brightening in $I_{\rm c}$, the class (C1), where the number of detected kernel pixels is around 1000, and finally a class (CP), which seems to exhibit a power-law dependence of the number of kernel points on the flare intensity. All three are indicated in Fig.~\ref{fig:classes}. 

\begin{figure}[!t]
\includegraphics[width=\textwidth]{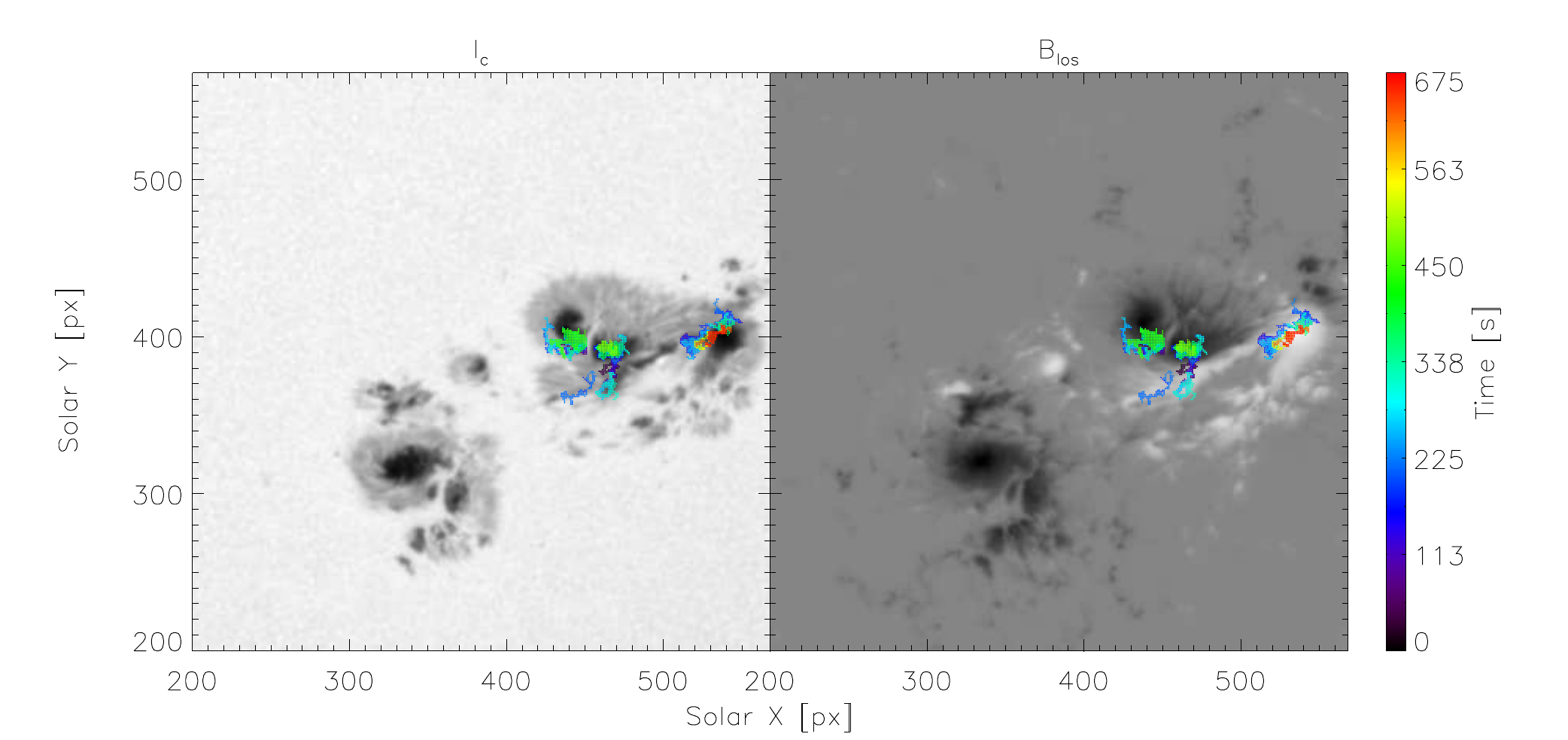}
\caption{Example of the flare belonging to CP class. This X2.2-class flare ignited in NOAA 11158 on 15 February 2011 with a maximum at 01:44 UT. On top of the background context images for SDO/HMI intensity (left) and line-of-sight magnetogram (right) the location of the flare kernels is overplotted. The colours represent the time stamp from violet (the beginning) to red (the end) colours. }
\label{fig:examples_nonflat}
\end{figure}

\begin{figure}[!t]
\includegraphics[width=\textwidth]{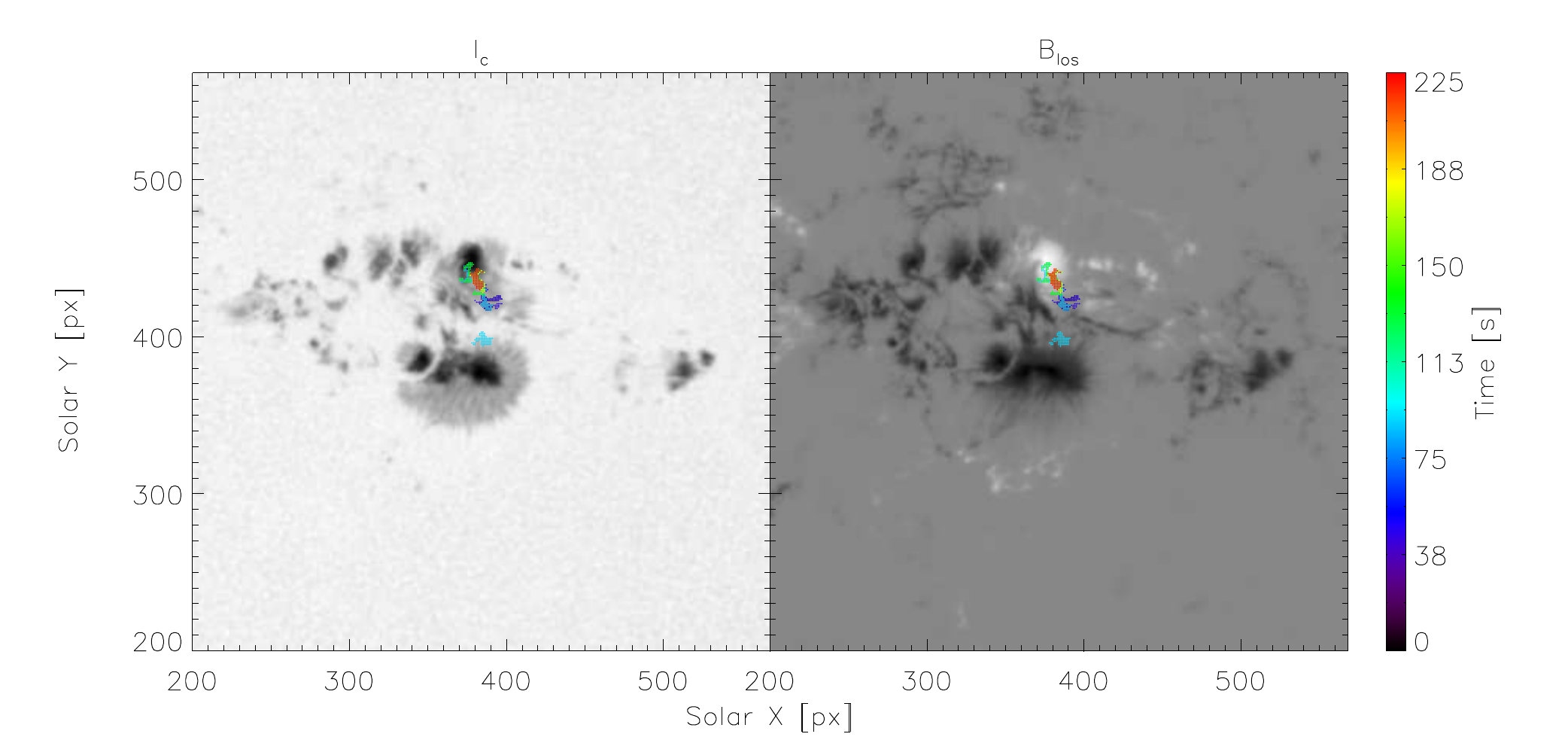}
\caption{Example of the flare belonging to C1 class. This M5.3-class flare ignited in NOAA 11283 on 9 June 2011 with a maximum at 01:35 UT. On top of the background context images for SDO/HMI intensity (left) and line-of-sight magnetogram (right) the location of the flare kernels is overplotted. The colours represent the time stamp from violet (the beginning) to red (the end) colours. }
\label{fig:examples_flat}
\end{figure}

Let us assume that there really are three classes of WLFs and let us find what makes those classes different. By plotting the flare kernels on top of the magnetogram we found that in the most cases, these point concentrate around strong polarity inversion lines. Generally, it is expected that the flare kernels in the lower atmosphere will reside in the footprints of quasi-separatrix layers \citep[QSL; see e.g. ][]{1995JGR...10023443P}, which is a generalisation of a too restrictive null-point configuration needed in an original formulation of the magnetic reconnection theory. To find the footprints of QSLs, we would need vector magnetic field measurements and/or extrapolation of the magnetic field configuration to the upper atmosphere, which would be beyond the scope of this simplistic study. Therefore we further studied only a gradient of the longitudinal magnetic field as a proxy for those regions by means of polarity inversion regions. \cite{1996A&A...308..643D} showed that the QSLs are usually found around polarity inversion lines. By eye we had a feeling that the CP flare kernels are located in regions with very strong magnetic field gradients, whereas C1 flares in regions, where the magnetic field gradient is not so strong and the field has often a diffuse character. Examples of active regions and detected flare kernels for CP and C1 flares are shown in Figs.~\ref{fig:examples_nonflat} and \ref{fig:examples_flat}, respectively. To confirm or reject the hypothesis, we investigated the value of the gradient of the longitudinal magnetic field $B_{\rm los}$, which is normalised by the maximum of the field in the field-of-view. Hence we studied the value
\begin{equation}
f(\bvec{r})=\frac{|\nabla B_{\rm los}(\bvec{r})|}{{\rm max}_{\bvec{r}} |\nabla B_{\rm los}(\bvec{r})|}, 
\end{equation}
where $\bvec{r}$ is a positional vector in the field-of-view. Value $f$ lies between 0 and 1, where 1 means that the given point of the flare kernel is exactly at the maximum gradient of the magnetic field in a given active region. 
\begin{figure}[!t]
\includegraphics[width=0.5\textwidth]{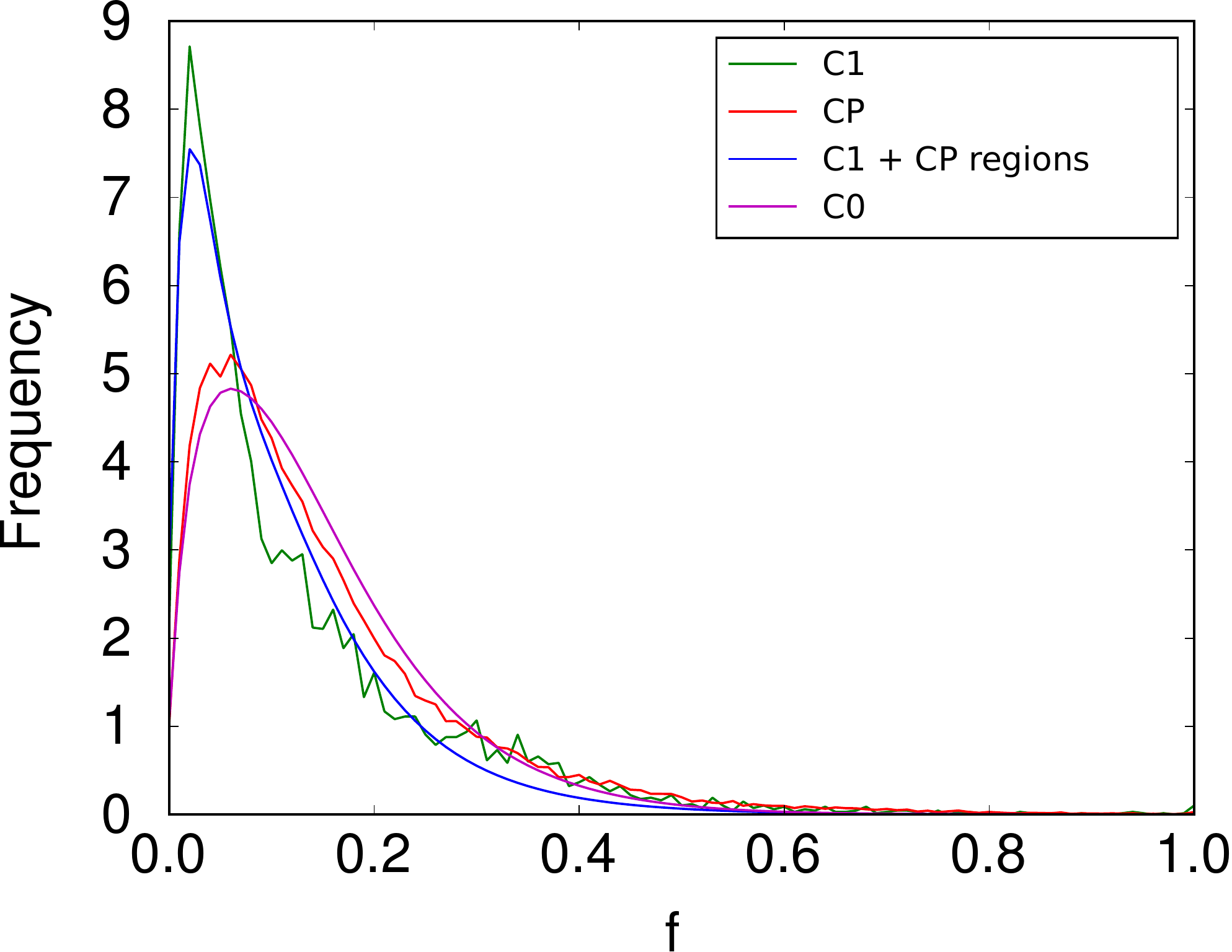} \\
\includegraphics[width=0.5\textwidth]{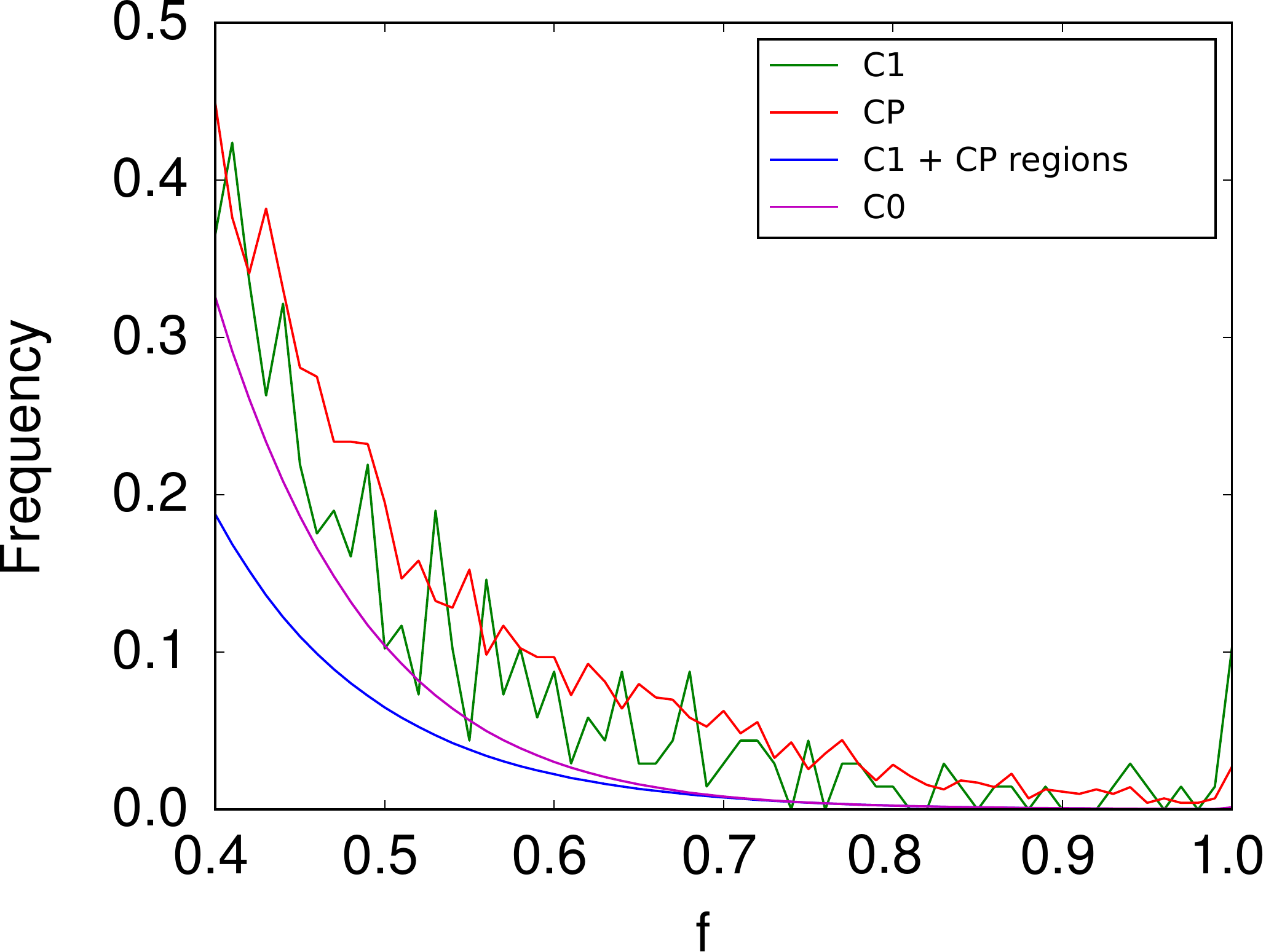}
\caption{Histograms of quantity $f$ for a set of situations in the full range of $f$ values (upper panel) and their magnification in the high-tail part (bottom panel).}
\label{fig:histograms}
\end{figure}

We investigated namely four following situations: 1. $f$ values in the regions without VL emission, 2. $f$ values in the regions with VL emission, and 3. and 4. $f$ values in the flare-kernel points for flare classes C1 and CP, respectively. The histograms (normalised so that their integral be unity) for these four situations are displayed in Fig.~\ref{fig:histograms}. Occurrence frequencies of $f$ in the case of C1 and CP flare kernels are larger than in the whole regions in the high-tail range, essentially meaning that the flare kernels do systematically appear in the regions of a large gradient of the magnetic field, where also the appearance of current sheets is expected. This finding is in agreement with models of magnetic reconnection in flares. In the high-tail range, the histograms of C1 and CP are not different, as the according to the Kolmogorov-Smirnof test, there is a 71\% probability that the apparent differences are due to chance. On the other hand C1 and CP histograms differ significantly in the low-values range, where for the CP-class flares the $f$ values seem to be significantly larger with a level of a statistical significance better than 99\%. 

It is not surprising that the histogram curves of $f$ in the C1 and CP regions lie below both curves in large-values range and in between in a smaller-values range, as it contains the information about the magnetic field gradients of both classes combined. On the other hand, the histogram of magnetic field gradients is somewhat surprising for the class C0, which seems quite similar as the histogram for CP class, despite the fact that according to the Kolmogorov-Smirnof test, they are strictly different. It might be that in C0 group the reason for non-detection of the VL emission is purely technical and that with a lower threshold in the code the VL kernel would be detected. In this case, however, the random noise would increase as well and such cases would have to be studied case by case. Although it certainly is possible to go through our small sample case by case, we aim at a much larger set of flares for the future and then the manual intervention is not thinkable.

We also need to point out that we deal with the $f$ value as if it is an exact value. One has to keep in mind that there are two kinds of errors appearing in $f$: first it is the measurement error of magnetic induction and its propagation through calculation of the derivative, and second the projection error due to the violation of the plan-parallel approximation. The uncertainty of the first kind is suppressed by smoothing of the magnetic field induction prior to the calculation of the derivatives. The uncertainties are further lowered in a statistical sense, because at least 7000 points in the case of C1 flares enter the histogram, ten times more in case of CP, and millions in the other two cases. The bin-to-bin oscillations seen in Fig.~\ref{fig:histograms} give an idea about the noise levels.

\section{Continuum or line-core emission? The case of X5.4-class flare}
Almost ending our report we have to discuss the origin of the emission in the SDO/HMI intensitygrams. It is not entirely certain that the $I_{\rm c}$ emission represents a white-light emission. $I_{\rm c}$ is a data product constructed from a set of filtergrams assuming a certain shape of the spectral line as \citep[from][]{2012SoPh..278..217C}
\begin{equation}
I_{\rm c} \equiv \frac16 \sum\limits_{j=0}^5 \left[ I_j + L_{\rm d} \exp\left(-\frac{(\lambda_j-\lambda_0)^2}{\sigma^2} \right)  \right],
\label{eq:ic}
\end{equation}
where $I_j$ are intensities in each of six filtergrams scanning the iron line with wavelength $\lambda_j$ and $\sigma$ is the line-width. 

When the shape of the line is different than assumed, e.g. when the line core is less deep e.g. due to the certain amount of emission in the core, reconstructed $I_{\rm c}$ may also be affected. To shed some light on this issue, we looked at line depth $L_{\rm d}$, another data products available from JSOC. 

For a short assessment of this issue we chose the case of the strongest flare in our sample, the X5.4-class flare that occurred on 6th March 2012. 
\begin{figure}[!t]
\includegraphics[width=0.7\textwidth]{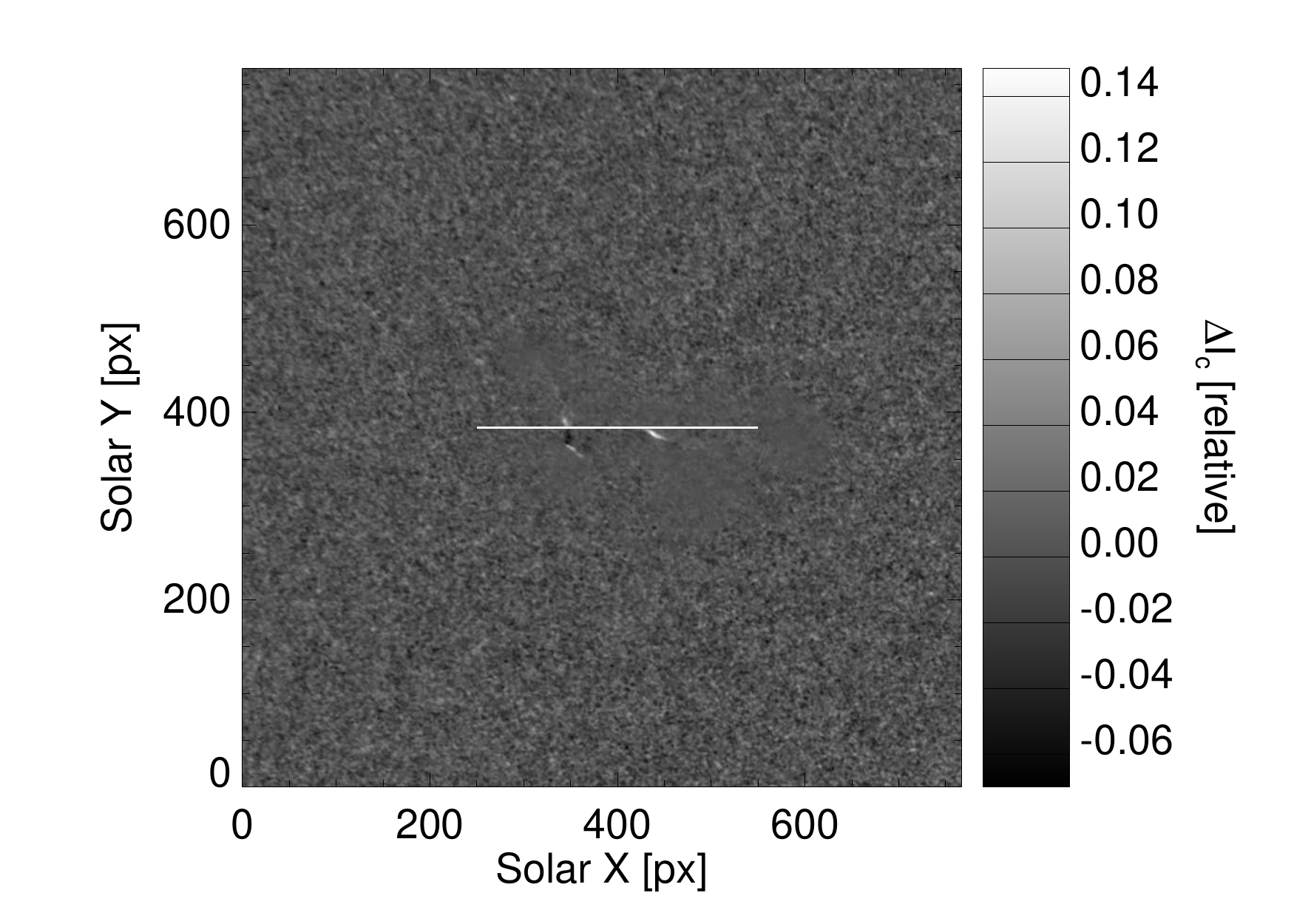}
\caption{A context frame of a running-differences movie at the time stamp of 7th March 2012 00:17:15 UT. The flare-kernel brightenings are well visible. The white line indicates the location for the time-distance diagrams in Fig.~\ref{fig:X54-time-distance}.}
\label{fig:X54-context}
\end{figure}

\begin{figure}[!t]
\includegraphics[width=0.49\textwidth]{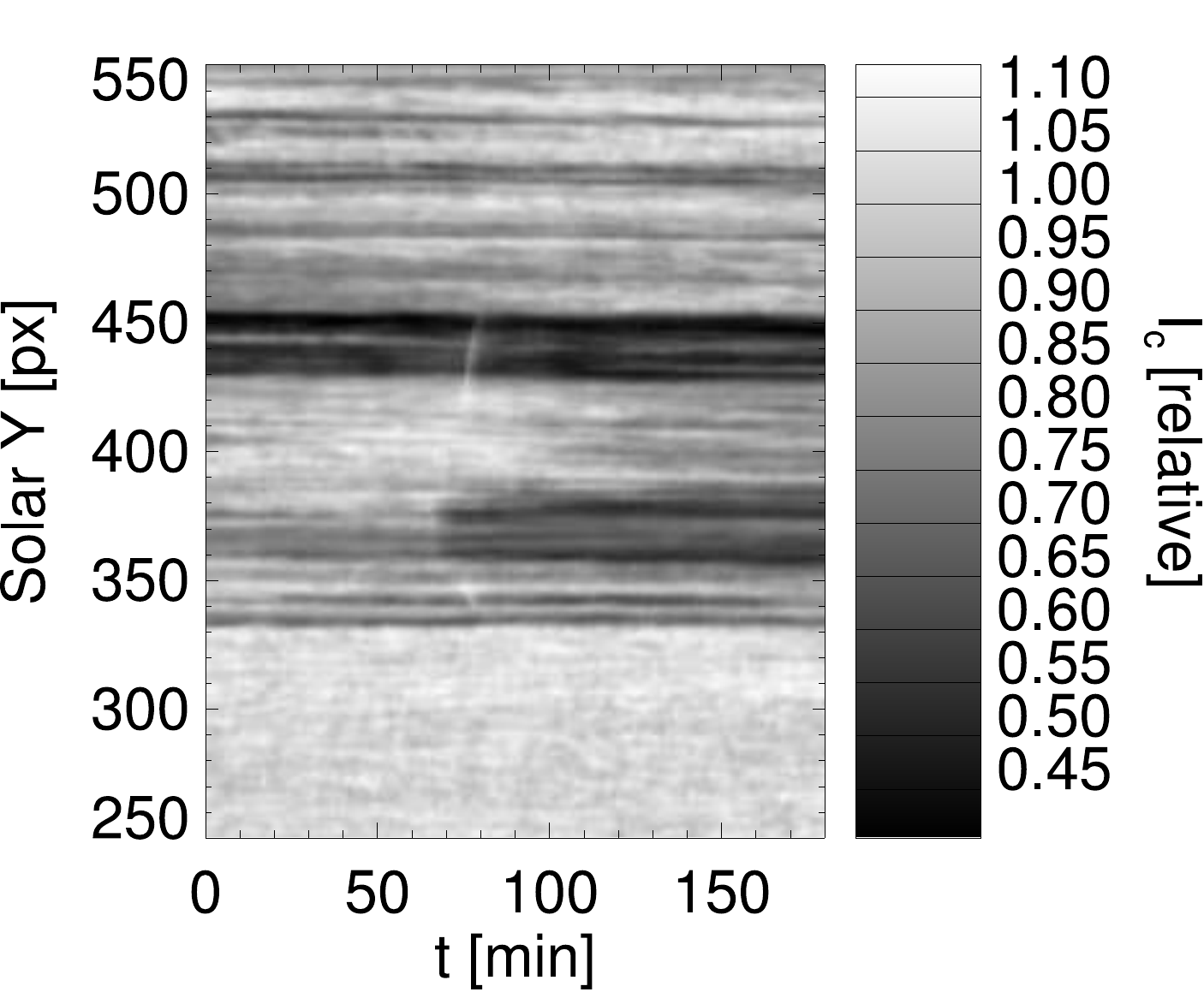}
\includegraphics[width=0.49\textwidth]{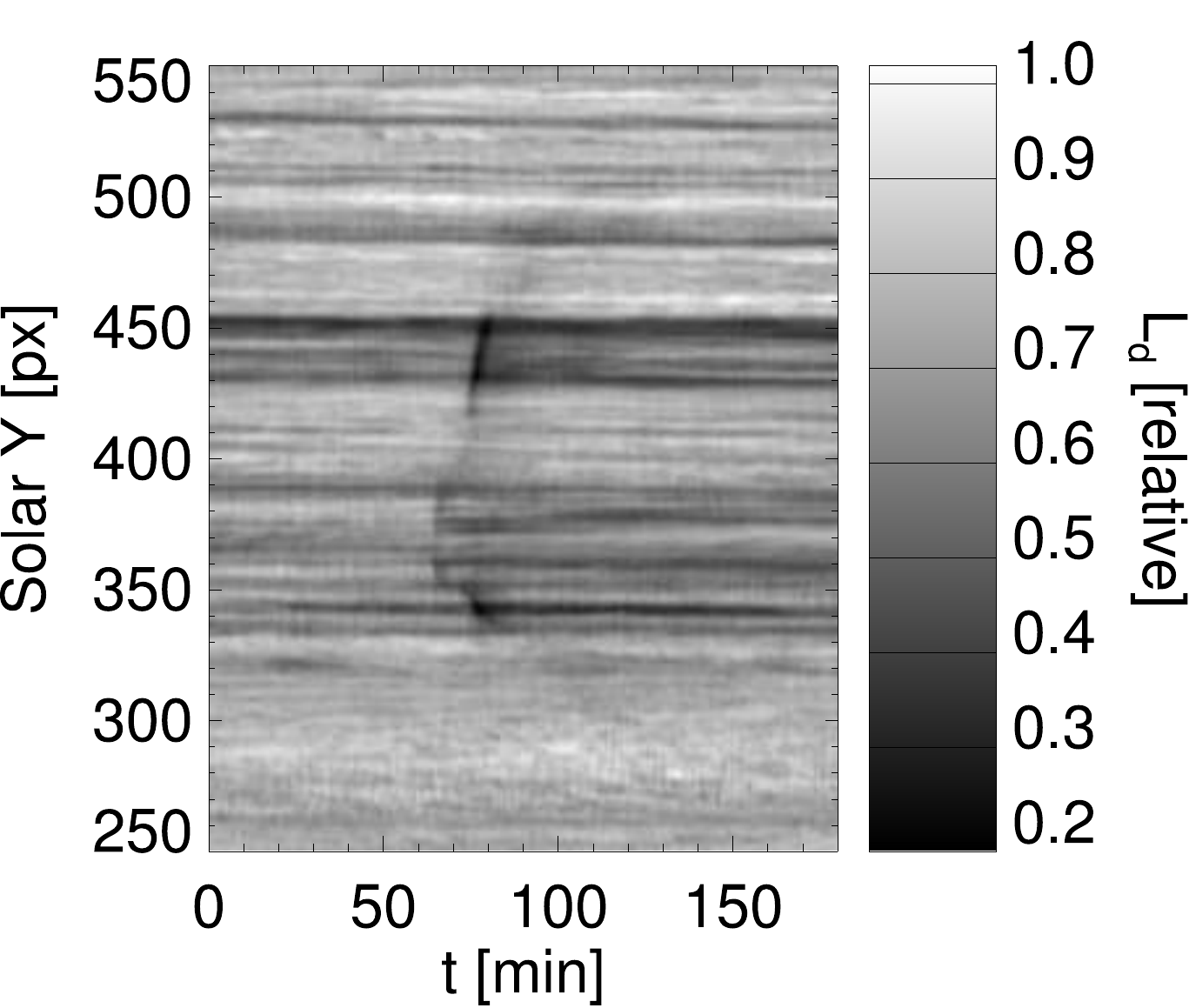}
\caption{Time-distance cuts along the line indicated in Fig.~\ref{fig:X54-context} for reconstructed continuum intensity $I_{\rm c}$ and line depth $L_{\rm d}$. The time starts on 6th March 2012 at 23:02:15 UT. The flare-kernel brightening connected to flare ribbons are propagating from time stamp of 65~min both up and down from $Y\sim 375$. Curiously, the $I_{\rm c}$ brightening do correlate with $L_{\rm d}$ darkenings. }
\label{fig:X54-time-distance}
\end{figure}

We downloaded all reconstructed data products available in JSOC and investigated the respective movies. The flare ribbons are well visible in $I_{\rm c}$, their visibility is enhanced when instead of $I_{\rm c}$ a running difference (the difference between consecutive frames) is displayed. An example of one frame is shown in Fig.~\ref{fig:X54-context}. Similar behaviour is seen in the movies of line-depth $L_{\rm d}$, only as darkenings, essentially meaning that the flare ribbons are connected with the shallower spectral line at the same place. 

A negative [one has to keep in mind that the ``orientation'' of $I_{\rm c}$ is opposite to $L_{\rm d}$, e.g., the shallower (lesser) $L_{\rm d}$ means a larger $I_{\rm c}$] correlation between $I_{\rm c}$ and $L_{\rm d}$ is apparently visible from (\ref{eq:ic}). By plotting cuts for various position in the field-of-view as a function of time, the deviations of both the $I_{\rm c}$ and $L_{\rm d}$ transients with respect to the secular trends may be estimated. We found that in most cases, the decrease of line-depth almost agrees (to within say 10\%) with the increase in continuum intensity. This issue deserves a future investigation by looking directly to the shape of the spectral line sampled by six band-pass filters of SDO/HMI, and we tested this idea using a simple model. 

\begin{figure}[!t]
\includegraphics[width=0.49\textwidth]{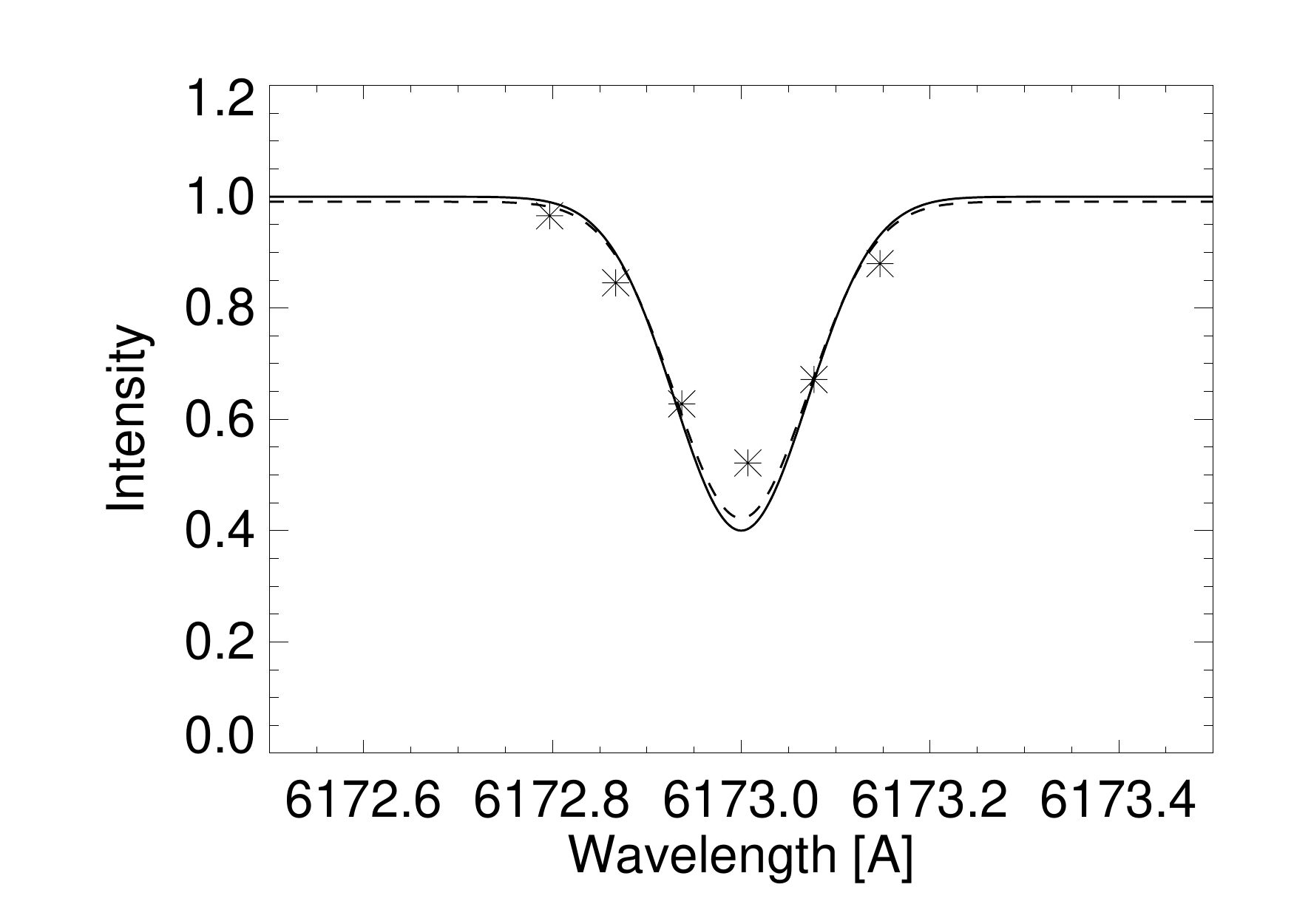}
\includegraphics[width=0.49\textwidth]{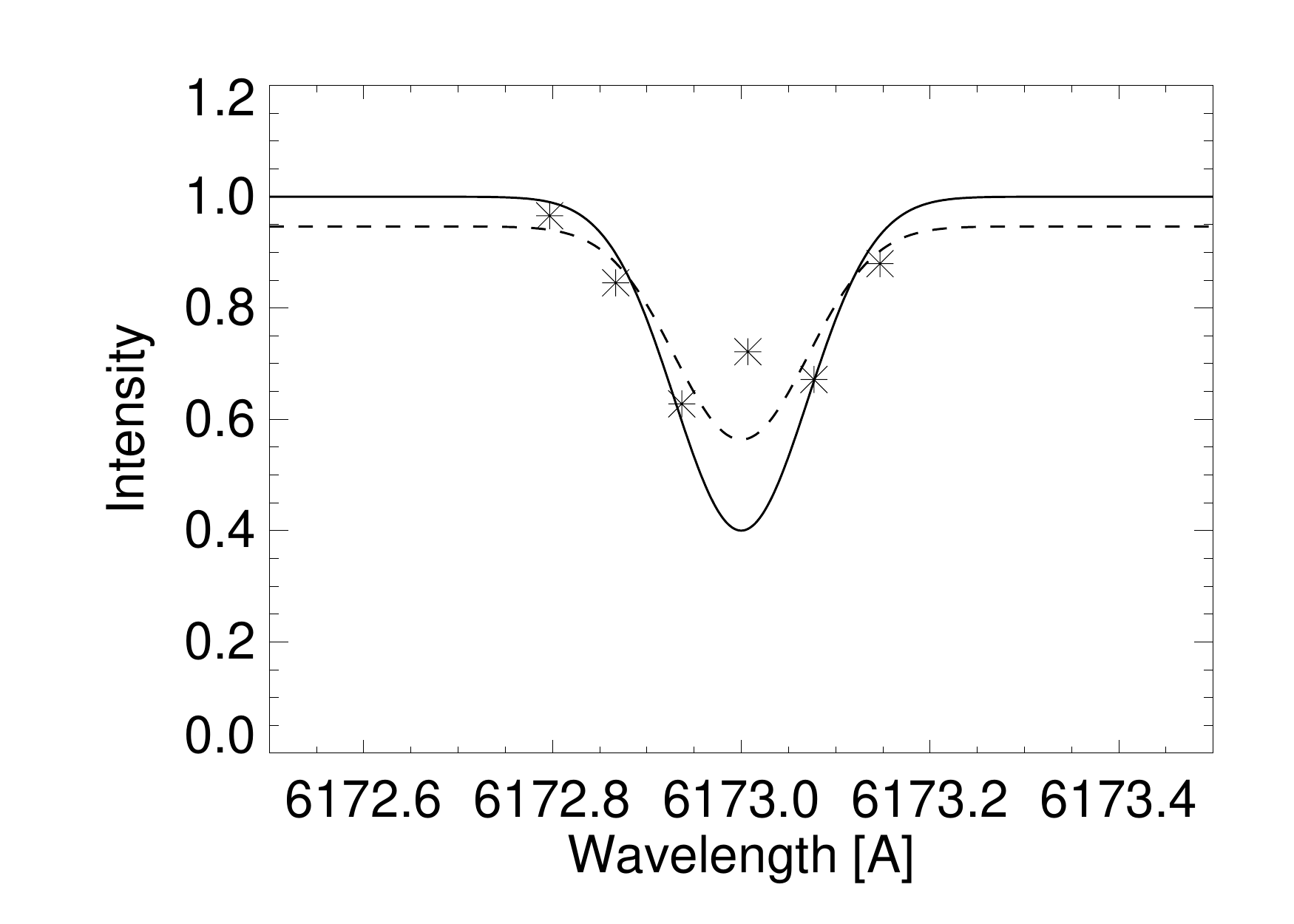}\\
\includegraphics[width=0.49\textwidth]{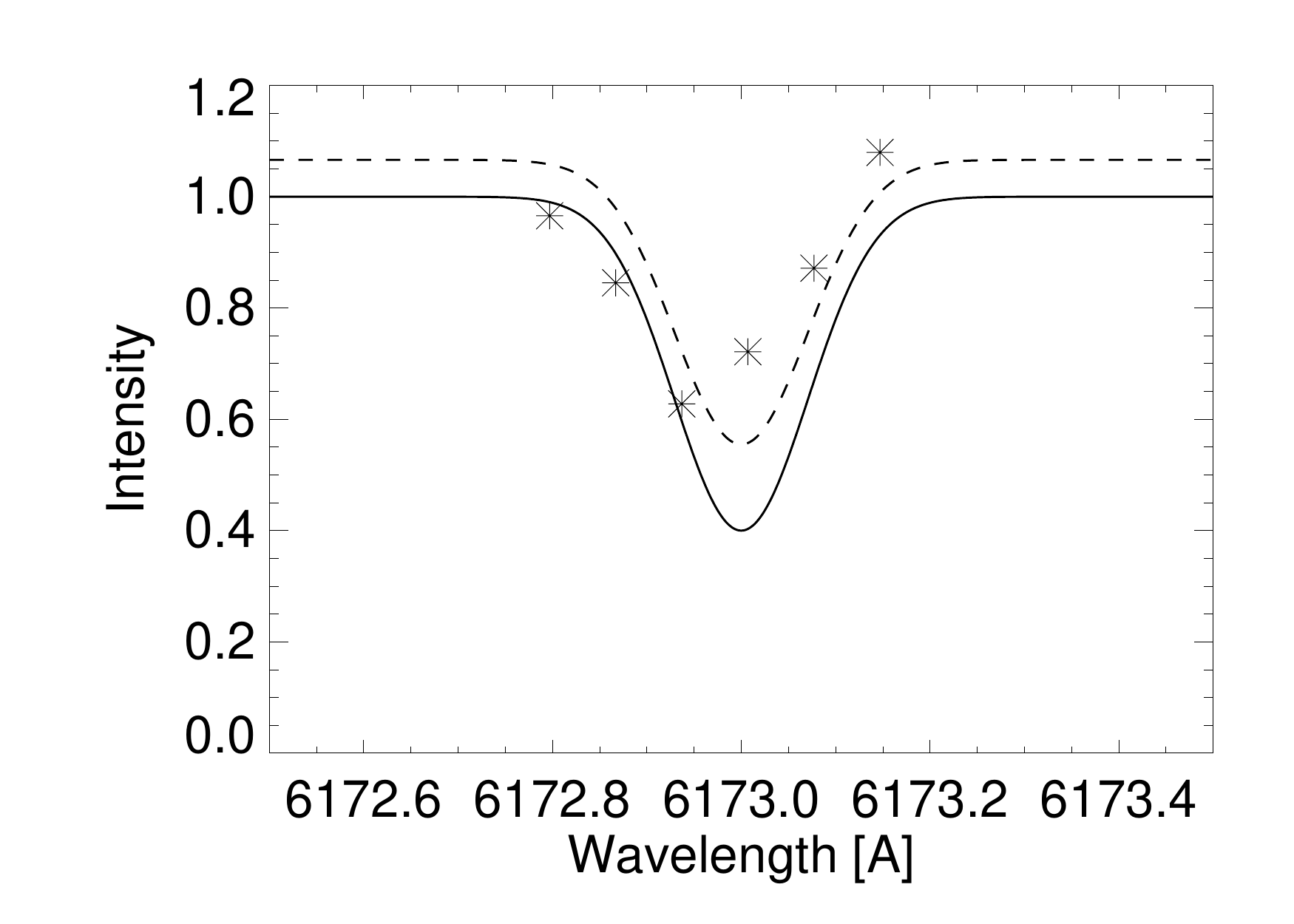}\\
\caption{Reconstruction of the spectral-line shape using the HMI algorithm in case of the quiet Sun (top left: $I_{\rm c}=0.98$, $L_{\rm d}=0.55$) and in case the line-core has an emission contribution (top right: $I_{\rm c}=0.93$, $L_{\rm d}=0.30$), and an asymmetry (bottom: $I_{\rm c}=1.07$, $L_{\rm d}=0.51$). The solid line represents the reference ``quiet-Sun'' spectral-line model, the points then the synthetic intensities obtained by applying the six idealised HMI band-pass filters to the modelled line profile, and the dashed line represents the reconstruction of the spectral-line shape from the six points. }
\label{fig:spline}
\end{figure}

We followed closely a recipe given by \cite{2012SoPh..278..217C}, where the means of derivation of HMI data products are given explicitly. We simulated two situations: first the quiet Sun situation and the situation, when the spectral line has a line-core emission. We then constructed six synthetic spectral-line scans with idealised HMI filters, and used the MDI-like algorithm \citep{2012SoPh..278..217C} to reconstruct the spectral line and corresponding data products, including the estimate of the continuum intensity $I_{\rm c}$ from (\ref{eq:ic}). This test is demonstrated in Fig.~\ref{fig:spline} in the upper row. It can be seen that in the case of the quiet Sun with an undisturbed spectral-line profile the reconstruction works well. However, when we introduced the artificial emission in the core of the line (other synthetic spectral scans were left unchanged), the reconstructed profile was shallower, with decreased line depth and also decreased continuum intensity. Thus the line-core emission itself cannot explain the apparent increase of the estimate of continuum intensity. 

Having a simple model at our disposal, by using a trial-and-error approach we searched for the situation, which will lead in what we observe: the increase of $I_{\rm c}$ and decrease of $L_{\rm d}$. We found that for instance a strong line asymmetry together with line-core enhancement (as seen e.g. in Fig.~\ref{fig:spline} bottom panel) produces such a behaviour. Without the line-core enhancement, the asymmetry itself does not produce a required line-depth decrease. Again, our simple test only indicates that the HMI brightening registered in flare kernels might be an artefact of the procedure, in which the HMI observables are produced. A careful investigation is needed by examining the individual filtergrams in detail.

\section{Concluding remarks}

We developed an automatic routine able to detect the signatures of WLF in the spatio-temporal datacubes composed from SDO/HMI intensitygrams. With a little effort, the code can be modified to be applied to any suitable datacube. We found optimal values of free parameters of the code; should the code be applied to other data set, a new tuning would have to be done. We found that the VL emission may be detected in flares as weak as M1.8, where this value is only coincidental and we assume that even weaker flares might depict a VL emission. In the future, we will extend our sample to include also weaker flares. 

For some flares there seems to be a power law connecting the flare intensity to the area of the VL kernels, whereas some seem to have the kernels confined to a certain limit. We found indications that for those two groups, there might be a difference in the configuration of the magnetic field, where the flares with a power-law dependence of the area of the kernels on the flare strength do seem to be observed in regions of very strong gradients of the magnetic field. It remains to be seen whether the two groups investigated in our study coincide to some extent with ``type I'' and ``type II'' WLFs as reported by \cite{1995A&AS..110...99F}.

There are a few issues connected to processing of this particular data set. It is not entirely certain that the $I_{\rm c}$ emission truly represents a white-light emission, our preliminary assessment suggests that the origin of the observer brightening might easily be an artefact of the data production, e.g. in the case when the line core is enhanced together with the spectral line being asymmetrical. By analysing one example we did not come to any firm conclusion, and we will return to this issue later. Also an investigation of real filtergrams that served to derive $I_{\rm c}$ proxy, especially those far in the spectral line wings, would improve the judgement whether the emission detected by our code is a real continuum or an artefact. We will take both approaches in the future. Similar discussion on ultraviolet slit-jaw images taken near the Mg lines on IRIS probe appeared recently in literature \citep{2017ApJ...837..160K}.

Another issue is that, given the emission we detect indeed is a continuum, where in the solar atmosphere (at which depth) this continuum forms. Some studies (e.g. Heinzel et al., in preparation) suggest that this continuum may be a Paschen continuum forming in the chromosphere at heights of a few hundreds kilometers above level of $\tau=1$. In that case the emission we detect does not come from the photosphere, where the ``true'' white-light emission should originate. In spite of all our observational material that was collected in the archives, the white-light flares are still far from being understood.

\section*{Acknowledgement}
This paper summarises the results of the BSc. thesis of LM under supervision of M\v{S} at Faculty of Mathematics and Physics, Charles University. M\v{S} acknowledges the support of the institute research project RVO:67985815 to Astronomical Institute of Czech Academy of Sciences. 

\section*{References}
% \bibliography{biblio}

\end{document}